\documentclass[intlimits,twoside,a4paper]{article}

\usepackage{graphicx}
\usepackage[cp1251]{inputenc}

\usepackage{cmpj3}

\issue{2017}{20}{3}{33603}
\doinumber{10.5488/CMP.20.33603}

\title[Water-DME properties from molecular dynamics simulations]
{Isobaric-isothermal molecular dynamics computer simulations of the properties of  
water-1,2-dimethoxyethane model mixtures} 

\author[J. Gujt, H. Dominguez, S. Sokolowski, O. Pizio]{J. Gujt\refaddr{label1}, H. Dominguez\refaddr{label2}, 
S. Sokolowski\refaddr{label3}, O. Pizio\refaddr{label2}\thanks{On sabbatical leave from 
Instituto de Quimica de la UNAM. 
Corresponding author: oapizio@gmail.com.}}

\addresses{
  \addr{label1}Chair of Theoretical Chemistry, Faculty of Chemistry,
  University of Duisburg-Essen, 
D-45141 Essen, Germany
\addr{label2} Instituto de Investigaciones en Materiales, Universidad Nacional Aut\'{o}noma de M\'{e}xico,
Circuito Exterior, 04510, Cd. de M\'{e}xico, M\'{e}xico
\addr{label3}Department for the Modelling of Physico-Chemical Processes, Maria Curie-Sklodowska University, \\
Lublin 20-614, Poland
}

\date{Received May 8, 2017, in final form August 4, 2017}

\begin{document}
\maketitle

\begin{abstract}
Isothermal-isobaric  molecular dynamics simulations 
have been performed to examine a broad set of properties 
of the model  water-1,2-dimethoxyethane (DME) mixture as a function of composition. 
The SPC-E and TIP4P-Ew water models and the modified TraPPE model for DME were applied.
Our principal focus was to explore the trends of behaviour of the structural
properties in terms of the radial distribution functions, coordination numbers and
number of hydrogen bonds between molecules of different species, and of conformations of DME
molecules. Thermodynamic properties, such as
density,  molar volume,  enthalpy of mixing and heat capacity at constant pressure
have been examined.  Finally, the self-diffusion
coefficients of species and the dielectric constant of the system were calculated
and analyzed.

\keywords  water-DME  mixtures, thermodynamic properties, 
self-diffusion coefficient, dielectric constant, molecular dynamics

\pacs 61.20.-p, 61.20-Gy, 61.20.Ja, 65.20.+w

\end{abstract}

\section{Introduction}
\label{1}

It is our honour to dedicate this manuscript to the memory of an
extraordinary scientist and extraordinary person Jean-Pierre  Badiali.  
One of us (O.P.), in particular, would like to appreciate friendship, 
scientific and non-scientific discussions with Jean-Pierre during his visits
in Ukraine and Mexico in past decades. J.-P. Badiali has made important
contributions in the area of theoretical electrochemistry and one of his
interests was in the properties of electrolyte solutions that
involve combined solvents with water and organic liquid 
components~\cite{badiali1,badiali2}.
Our present contribution is on this topic. 

Various experimental techniques have been applied to several water-organic solvent mixtures
for specific purposes. As a result, numerous valuable experimental data have been accumulated.
Still, even for most popular systems of experimental interest such as water-alcohols,
water-dimethylsulfoxide (DMSO), water-amides, these studies have not been 
entirely comprehensive. Consequently, computer simulations, most frequently the molecular dynamics
studies, have been performed aiming at explaining experimental findings and 
accumulation of data for the properties difficult or impossible to access in a laboratory.
Moreover, computer simulations permit to discern different factors contributing to a given
result and can provide a profound understanding of the trends of the behaviour of the properties
of interest. 

For specific purposes of our project, it is worth mentioning that molecular dynamics computer
simulations have been applied to mixtures of water with different organic solvents in a large
number of works. We cite only some of them, just to provide the reader a taste 
of the objectives and
principal results for a few selected co-solvents. Namely, water mixed with different alcohols
was studied in, e.g., \cite{wensink,pusztai1,galicia1,galicia2,patey1,patey2,patey3,bopp1}, 
see references therein as well. On the other hand, liquid mixtures of water with 
dimethylsulfoxide (DMSO) have been explored in~\cite{vaisman,luzar3,gunsteren1,jedlovszky1,samios,gunsteren2,skaf1,skaf2,vishnyakov,mancera,bagchi1,jedlovszky2,bagchi2,perera,jedlovszky3,gujt}.
Less attention has been paid to the exploration of the behaviour of water-dymethylformamide (DMF) 
mixtures~\cite{bopp2,samios1,zoranic1,zoranic2,takamuku1,samios2,samios3}.

Finally, liquid mixtures of water with 1,2-dimethoxyetane (DME or monoglyme) were studied by 
molecular dynamics computer simulations in, e.g., \cite{siepmann,sadowski,bedrov,roccatano}.
Meanwhile, DMSO and DMF are widely used as solvents and reaction media in laboratory studies,
the DME molecule is of interest as it is the smallest element of the polyoxyethylene (POE),
the water soluble polymer with various applications in biomedicine.
The hydrophilic and hydrophobic groups are combined in the DME molecular structure of the form 
CH$_3$-O-CH$_2$-CH$_2$-O-CH$_3$, the intra- and intermolecular interactions are such that the DME is
highly soluble in water and in other solvents. Significant conformational changes of the
intramolecular structure depend on the co-solvent and are of interest to explore in the
perspective of other chemical and biochemical systems. 

In close similarity to various previous studies, this work is performed in the framework 
of isobaric-isothermal molecular dynamics computer simulations.
The structure of water-DME mixtures on composition is explored in terms of different descriptors,
namely the radial distribution functions and coordination numbers, hydrogen bonding statistics
and population of most abundant conformations of DME species. Thermodynamic aspects of
the behaviour of mixtures in question are discussed in terms of mixing properties. 
Insights into the dynamic properties are obtained by exploring the mean square displacements 
as a  function of time and the resulting self-diffusion coefficients of species. 
Finally, we  discuss the behaviour of the dielectric constant of the solutions upon changes 
in the solvent composition. 
Our study is restricted to room temperature and ambient pressure 
(1~bar), only  the chemical composition of the mixed solvent represent the explored variable.
Wider insights into thermodynamic properties, analyses of an broad set of dynamic and 
dielectric properties, would require additional work.

\section{Model and simulation details}

Our calculations have been performed in the isothermal-isobaric ($NPT$) ensemble 
at 1~bar, and at a temperature of 298.15~K. We used the GROMACS 
package~\cite{gromacs} 
version 4.6.5.
Solely the modified TraPPE united atom model (TraPPE-UA) for
DME~\cite{sadowski} was used in our calculations. It consists of six interaction 
sites (CH$_3$, O, CH$_2$, CH$_2$, O, CH$_3$).
All the interaction parameters (charges, $\sigma$ and $\epsilon$ for 
Lennard-Jones interactions), 
harmonic bonds lengths, bond strengths, bend and torsional angles were taken from the
table~1 of~\cite{sadowski}. Most important is that the model is perfectly well defined. 
In this aspect, we would like to 
repeat the comment by Fischer et al.~\cite{sadowski} ``we did not try to reproduce their data, because
the employed force field parameters they used are not clearly indicated'' referring to the model
of Bedrov et al.~\cite{bedrov2,bedrov}. On the other hand, we have failed to reproduce the data
for the DME united atom type model given in \cite{roccatano} for the reason just mentioned and 
due to misprints in table~1 of this article.

For water in the present study, the SPC-E model~\cite{spce}
and the TIP4P-Ew model~\cite{horn} were used. The Lorentz-Berthelot combination rules 
for diameters and energies were used to determine 
the cross parameters. Nevertheless, we checked the effect of a combination rule on mixture
density as function of composition by applying the geometric combination rule (rule~3 in GROMACS
nomenclature). 
The nonbonded interactions were cut off at 1.59~nm, similar to \cite{roccatano},
 and the long-range electrostatic interactions were handled by the
particle mesh Ewald method implemented in the GROMACS software package (fourth 
order, Fourier spacing equal to 0.12). 
The van der Waals tail correction terms to the energy and pressure were taken into account.
In order to maintain the geometry of the water and DME molecules, the LINCS 
algorithm was used.

For each system, a periodic cubic simulation box was set up. 
The GROMACS genbox tool was employed to randomly place all particles into the 
simulation box. The total number of molecules  was kept fixed at 3000. 
The composition of the mixture is described by the mole fraction of DME molecules, 
$X_{\text{dme}}$, $X_{\text{dme}}=N_{\text{dme}}/(N_{\text{dme}}+N_{\text w})$. 

To remove possible overlaps of particles introduced by the procedure of 
preparation
of the initial configuration, each system underwent energy
minimization using the steepest descent algorithm implemented in the GROMACS
package. Minimization was followed by a 50~ps $NPT$ equilibration run at 298.15~K and 1~bar using a timestep of 0.25~fs. 
We used the Berendsen thermostat and  barostat with $\tau_T = 1$~ps and $\tau_P = 1$~ps 
during equilibration. Constant value of $4.5 \cdot 10^{-5}$~bar$^{-1}$ for the compressibility of the mixtures was 
employed.
 
The V-rescale thermostat and Parrinello-Rahman
barostat with $\tau_T = 0.5$~ps and $\tau_P = 2.0$~ps
and the time step 2~fs were used during production runs. To test
this thermostat and barostat, we have obtained 88.5~J/mol$\cdot$K for the heat 
capacity of the SPC/E model. This value is close to 86.6~J/mol$\cdot$K reported by Vega et 
al.~\cite{vega} (experimental 
result is 75.3~J/mol$\cdot$K). On the other hand, we obtained 191.2~J/mol$\cdot$K for 
the heat capacity of DME, the experimental value reported by Trejo et al.~\cite{trejo} 
is 191.14~J/mol$\cdot$K. Moreover, the DME molar volume coming out from our calculations is 
105.2~cm$^3$/mol favourably comparing to the experimental result 104.54~cm$^3$/mol.
Berendsen type control of temperature and pressure is not satisfactory in this aspect.

Statistics for each mole fraction for some of the properties were collected over
several 10~ns $NPT$ runs, each started from the last configuration of the 
preceding run. The total trajectory was not shorter than 60~ns. Actually, the heat capacity 
and the dielectric constant are the most demanding properties. 

\section{Results and discussion}

\subsection{Density and mixing properties}

To begin with, we would like to discuss the dependence of density of the
water-DME mixture as function of its composition.  Previous comparisons, 
see figure~5 of \cite{sadowski}, were made at $T = 318$~K using TIP4P-Ew water
model and several models for DME. It has been concluded that the combination 
of TIP4P-Ew with two all-atom type models~\cite{anderson} performs worse than 
the the combination of TIP4P-Ew with all-atom type model of \cite{bedrov2}
and the united atom modified TraPPE model of~\cite{sadowski}, in comparison with 
the experimental data from \cite{hazra}.  
On the other hand, a comparison of the theoretical and experimental data given in figure~1 of
\cite{roccatano} again refers to $T = 318$~K, but involves the SPC model of water~\cite{hermans}.
The quality of theoretical predictions is satisfactory.

\begin{figure}[!t]
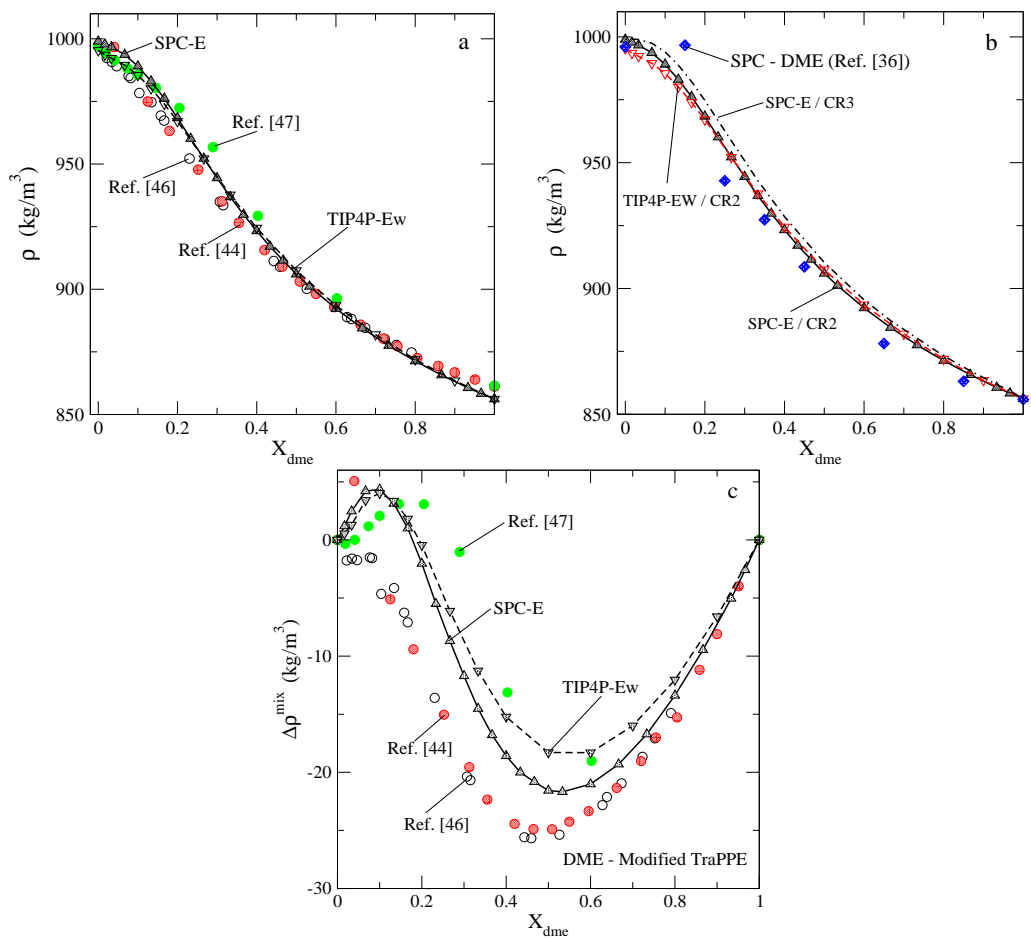

\begin{center}
\includegraphics[width=6.5cm,clip]{fig1a.eps} \quad
\includegraphics[width=6.5cm,clip]{fig1b.eps}\\
\includegraphics[width=6.5cm,clip]{fig1c.eps}
\end{center}
\vspace{-5mm}
\caption{\label{fig1}(Color online) Panel (a): Composition dependence of the density of 
water-DME mixtures 
from constant pressure-constant temperature simulations ($T = 298.15$~K, $P = 1$~bar)
in comparison with the experimental data from~\cite{hazra,renard,marchetti}.
Panel (b): Simulation results for different models and from different combination rules.
Panel (c): Excess mixing density on composition.} 
\end{figure}

Results of our calculations at $T = 298.15$~K are given in two panels, (a) and (b), of figure~\ref{fig1}.
In order to perform comparisons, we have used three sets of experimental data, namely
from \cite{hazra,renard,marchetti}. From the simulation data in panel~(a) we learn that
the theoretical curves slightly overestimate the density in water-rich region comparing with
the results of \cite{hazra,renard} and slightly underestimate the density in
the DME-rich region, but this discrepancy is in fact rather small. Both models, i.e.,
the SPC-E and TIP4P-Ew, are equally good. In addition, the application of geometric 
combining   rule does not yield improvement of the dependence of density on composition. 
It can be seen that the application of another DME model in combination with the SPC water model
(the data were compiled from table~IIS of \cite{roccatano}) yields less accurate results.

It is of interest to explore the accuracy of performance of the models 
by studying deviations from the ideal type behaviour of different properties. 
One test for the species modelling is illustrated by the calculations of the excess mixing
density,  $\Delta \rho^{\text{mix}} = \rho-(1-X_{\text{dme}})\rho_{\text{water}}-X_{\text{dme}}\rho_{\text{dme}}$
(similar type of an expression is used below to evaluate other mixing properties).

Our results are shown in panel~(c) of figure~\ref{fig1}.  
According to the experimental data of \cite{hazra,renard}, negative deviation from the ideal
type of  mixing is observed in the
entire range of composition whereas a weakly pronounced positive deviation
in the water-rich region comes out from the experimental data of~\cite{marchetti}. 
The simulation curves yield a reasonable description of $\Delta\rho (X_{\text{dme}})$ behaviour. 
The SPC-E water-DME model is closer to
the experimental predictions, comparing to the TIP4P-Ew one. The most pronounced deviation 
is observed in the interval of compositions between 0.5 and 0.6 whereas the experimental 
predictions of \cite{hazra,renard} show a maximum deviation from the ideal type of behaviour 
of this property at $X_{\text{dme}} \approx 0.45$. 
It is worth mentioning that for mixtures of water with methanol, ethanol and 1-propanol,
as well as for water-DMSO mixtures, $\Delta \rho^{\text{mix}}$ is positive in the entire
chemical composition range, see, e.g., figure~2 of \cite{wensink} and figure~1 of
\cite{gujt}, respectively. In the present system, the behavior of  $\Delta \rho^{\text{mix}}$ 
is different due to the intrinsic shape of the DMF molecule and due to the way 
these molecules accommodate one another.

\begin{figure}[!t]
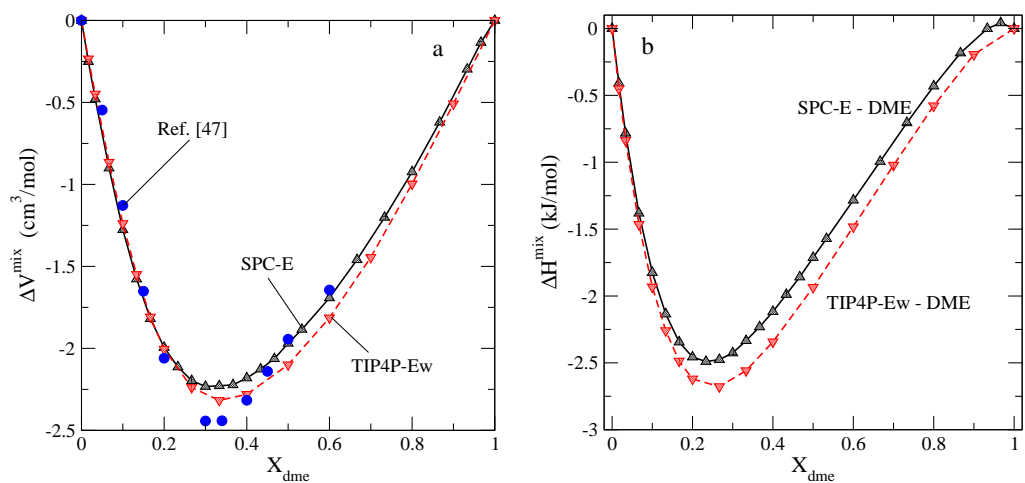

\begin{center}
\includegraphics[width=6.5cm,clip]{fig2a.eps}\quad
\includegraphics[width=6.5cm,clip]{fig2b.eps}
\end{center}
\caption{\label{fig2} (Color online) Excess mixing molar volume and excess mixing enthalpy of water-DME mixtures
on composition.}
\end{figure}
\begin{figure}[!b]
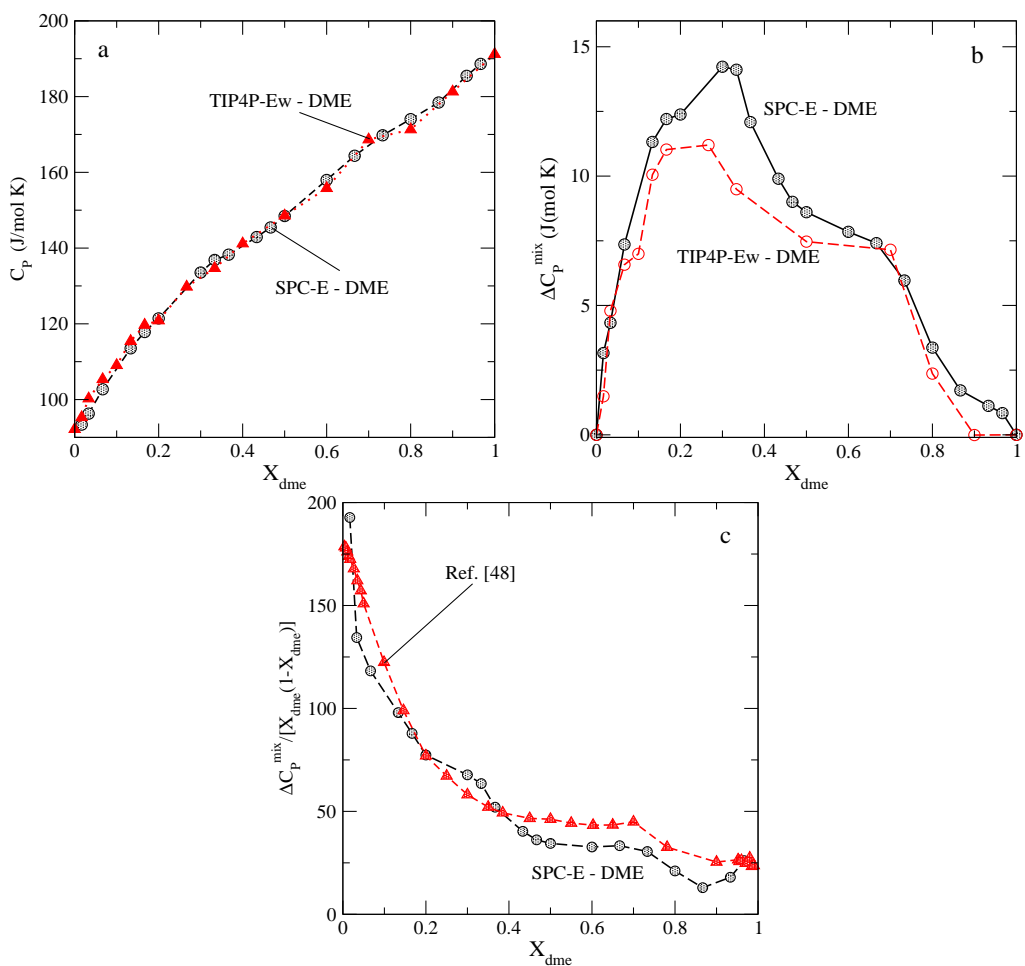

\vspace{-5mm}
\begin{center}
\includegraphics[width=6.5cm,clip]{add_fig2.eps}\quad
\includegraphics[width=6.5cm,clip]{fig2c.eps}\\
\includegraphics[width=6.5cm,clip]{fig2d.eps}
\end{center}
\caption{\label{fig3} (Color online) Panel (a): Molar heat capacity on mixture composition. 
Panels~(b) and (c): Excess mixing heat capacity of water-DME mixtures on composition.}
\end{figure}

Insights into mixing of species in the mixtures upon composition are commonly discussed
by inspecting a set of results given in figures~\ref{fig2} and \ref{fig3}. Namely, in figure~\ref{fig2}~(a) we present 
our results
for the mixing volume for two models for water combined with the modified TraPPE model for DME. They
are compared with the experimental results from \cite{marchetti}. Both sets of theoretical
results  rather well agree with experimental data. In particular, the simulations predict the magnitude
of volume contraction and the position of a minimum of $\Delta V^{\text{mix}}(X_{\text{dme}})$ at 
$X_{\text{dme}} \approx 0.3$.  
On the other hand, the energetic aspects of mixing at a constant pressure are given by the excess mixing
enthalpy, panel~(b) of figure~\ref{fig2}. Two versions for water-DME models predict maximal mixing trends around
$X_{\text{dme}} \approx 0.25$, i.e., almost at the same composition as the excess mixing volume. 
Unfortunately, we did not find experimental data to evaluate the precision of the simulation results.

Thermodynamic aspects of mixing can be interpreted in terms of the behaviour of molar heat capacity.
This property was not discussed in previous publications on water-DME mixtures.
The dependence of molar heat capacity on the composition coming from our simulations is shown in
panel~(a) of figure~\ref{fig3}. 
The starting point for pure water and the final point for pure DME of the
curve agree rather well with experimental predictions as we have already mentioned in the previous
section. Interestingly, the results were extracted just using GROMACS software 
without taking into account the quantum corrections.
The only interesting observation is that the heat capacity grows faster with $X_{\text{dme}}$ in 
the water-rich region and then starting from $X_{\text{dme}} \approx 0.3$ it changes almost linearly with
an increasing concentration of organic species. On the other hand, the excess mixing molar heat capacity
exhibits a maximum within this composition interval. Moreover, this peculiarity 
in $\Delta C^{\text{mix}}_P(X_{\text{dme}})$
coincides along $X_{\text{dme}}$ axis with the extrema for the excess mixing volume and enthalpy in figure~\ref{fig2}.
The accuracy of simulation predictions can be deduced from comparison with the experimental data
selected from a big set of data given in \cite{couture}. It appears that the simulation 
predictions agree with experiments reasonably well.
We would like to conclude the discussion of thermodynamic features of mixing just mentioning that
a more elaborate exploration of other fluctuation based thermodynamic properties, related 
to the experimental observations provided in \cite{george},  will
be given in a separate work.

\subsection{Pair distribution functions, coordination numbers and hydrogen bonding}

The microscopic structure of mixtures is usually and conveniently described in terms of 
various pair distribution functions. However, only very limited insights in this aspect 
were given in previous works. Specifically, only the pair distribution function, $g_{ij}(r)$,
describing the evolution of configurations of OW-O$_{\text{dme}}$ atoms upon changing the chemical 
composition in terms of $X_{\text{dme}}$ was given in figure~6 of \cite{sadowski} at $T = 298$~K, 
the evolution of OW-OW function was shown in figure~2S of \cite{roccatano} at $T = 318$~K.  
Various combinations of models for water and DME have been involved. Unfortunately, the 
structure factors from either X-ray or neutron diffraction experiments are not available in the
literature, to our best knowledge. Hence, the critical evaluation of the quality of simulation 
predictions similar to, e.g., \cite{galicia1} cannot be performed for the moment. 
Therefore, we restrict ourselves to a brief discussion of the trends of the behaviour of the microscopic
structure in this work.

\begin{figure}[!t]
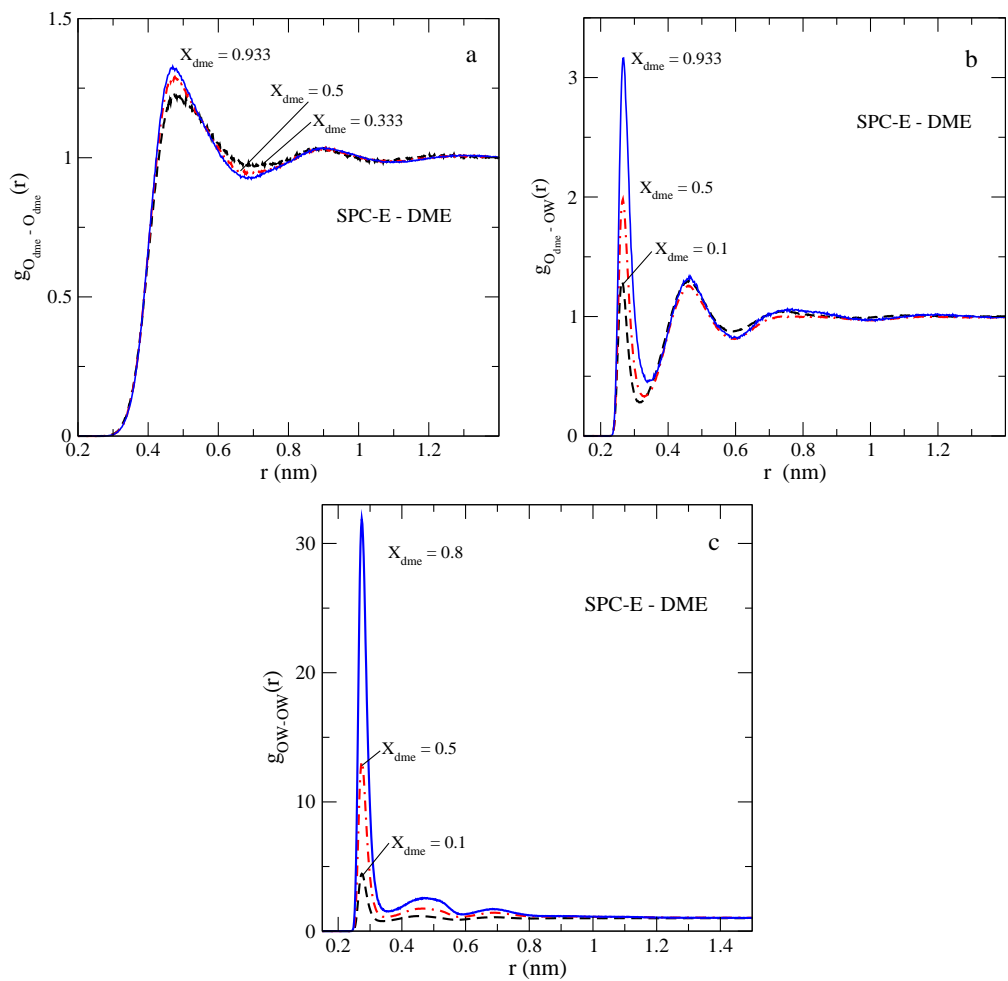

\begin{center}
\includegraphics[width=6.5cm,clip]{fig3a.eps}\quad
\includegraphics[width=6.3cm,clip]{fig3b.eps}\\ \vspace{2mm}
\includegraphics[width=6.5cm,clip]{fig3c.eps}
\end{center}
\vspace{-3mm}
\caption{\label{fig4} (Color online) Evolution of the pair distribution functions,  O$_{\text{dme}}$-O$_{\text{dme}}$, 
O$_{\text{dme}}$-OW and OW-OW with
changing solvent composition.}
\end{figure}

\begin{figure}[!t]
\begin{center}
\raisebox{0.1\height}{\includegraphics[width=5.5cm,clip]{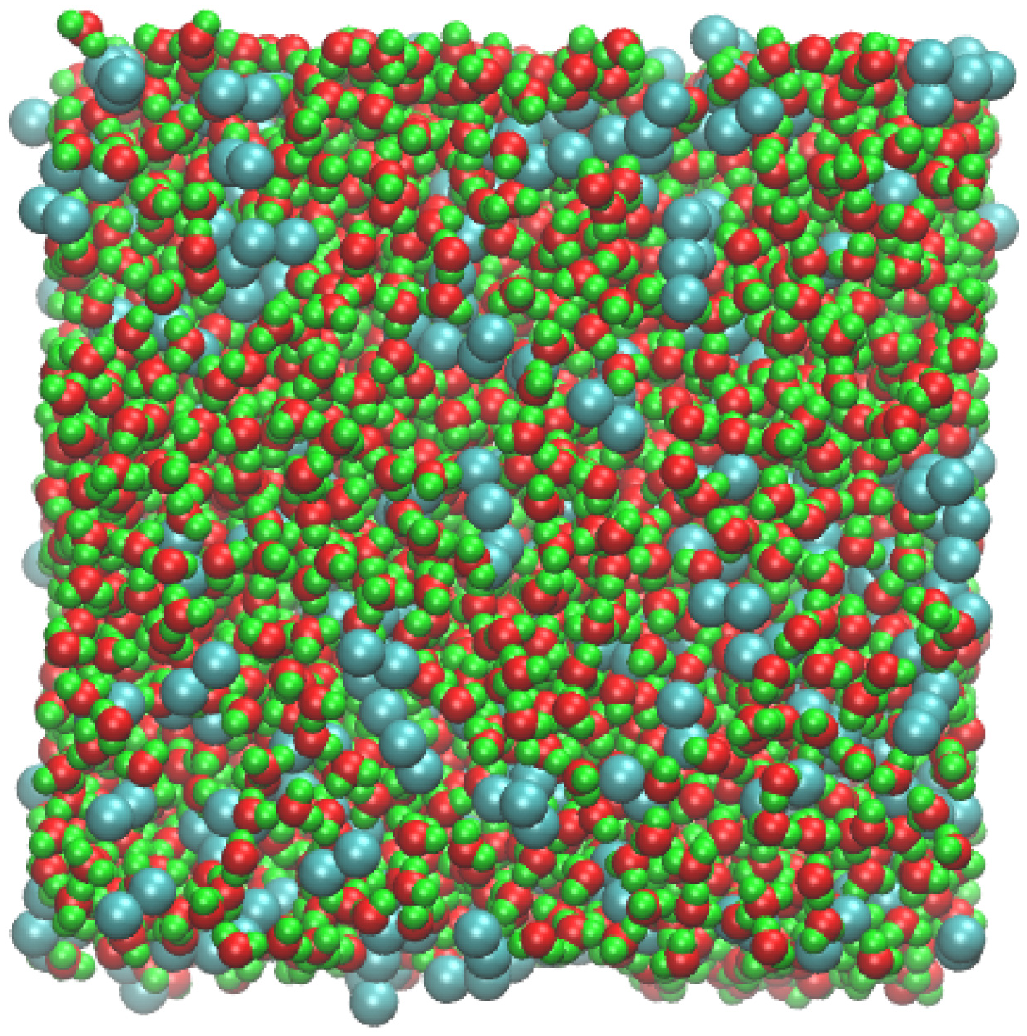}}\quad
\includegraphics[width=6.4cm,clip]{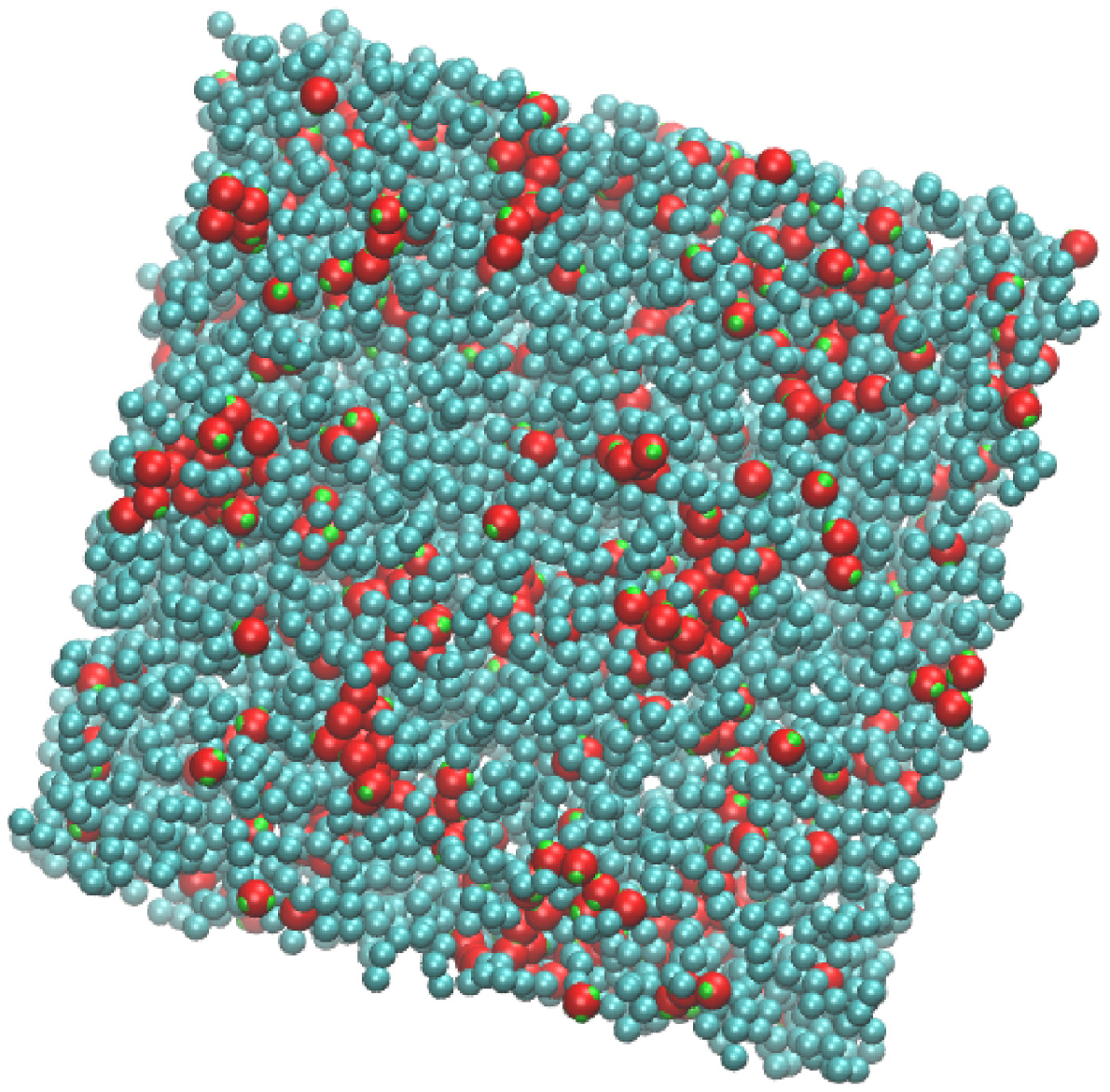}
\end{center}
\caption{\label{fig5} (Color online) Visualization of the distribution of water and DME species
in the simulation box for two values of mixture composition. Left panel is for
$X_{\text{dme}} = 0.1$ whereas the right panel is at $X_{\text{dme}} = 0.8$. Only the carbon sites
are shown for DME (cyan), water oxygens and hydrogens are given by red and green
spheres, respectively.}
\end{figure} 

Our first important observation is that the intermolecular pair distribution function, O$_{\text{dme}}$-O$_{\text{dme}}$,
is rather inert or, in other words, exhibits very small changes in a wide interval of composition,
panel~(a) of figure~\ref{fig4}. The second observation concerns the OW-O$_{\text{dme}}$ distribution function, panel~(b)
of figure~\ref{fig4}. Computer simulations predict that this distribution function is sensitive to the value of
$X_{\text{dme}}$ but only at small interparticle separations, i.e., close to their mutual contact. At larger
distances, we do not observe substantial changes of the shape of this distribution function.
Thus, the structuring of water around ether oxygens increases locally with increasing $X_{\text{dme}}$, 
miscibility of water and DME is actually expected. The most drastic changes can be seen for the
OW-OW distribution, panel~(c) of figure~\ref{fig4}. The first maximum of water distribution strongly increases
with increasing $X_{\text{dme}}$, i.e., when water fraction decreases. The second and third maxima of water 
distribution grow as well. However, the asymptotic value at unity is clearly seen. In other words,
the water density becomes locally heterogeneous.
However, the shape of OW-OW distribution does not suggest the formation of large water clusters.
These trends are qualitatively similar to what was observed for another DME model combined with the
SPC water model in \cite{roccatano}.
The observed  behaviour combined with the trends of OW-O$_{\text{dme}}$ changes, can indicate a certain tendency to
the development of  local
heterogeneity and even to demixing at a local scale for a particular combination of the
models of each species, as already discussed in~\cite{sadowski}. This is an important observation,
because it can have implications in the future  studies of solutions of salts and/or 
complex organic molecules in such combined solvents. In order to provide a better insight into
the distribution of particles of each species, we present visualization in two panels of figure~\ref{fig5}.
From the left-hand panel of this figure we learn that the DME molecules are rather uniformly distributed
in the medium with a predominant number of water molecules, cf., the pair distribution function 
at $X_{\text{dme}} = 0.1$ in figure~\ref{fig4}~(c). The O$_{\text{dme}}$-O$_{\text{dme}}$ distribution is of the type shown in 
figure~\ref{fig4}~(a). The right-hand panel of figure~\ref{fig5} describes the situation at $X_{\text{dme}} = 0.8$, cf. OW-OW 
distribution with very high first maximum and two well pronounced maxima at $r \approx 0.5$~nm
and at $\approx 0.7$~nm. The system is undoubtedly macroscopically homogeneous. However, 
the inspection of water molecules distribution leads to a conclusion that local heterogeneities
are present, the presence of associates of a few water molecules is quite probable.  
It is worth noting at this point that a detailed discussion of microheterogeneities
in aqueous amide mixtures was given in \cite{zoranic2}. We are unaware of
similar developments for water-DME mixtures. 
 
\begin{figure}[!t]
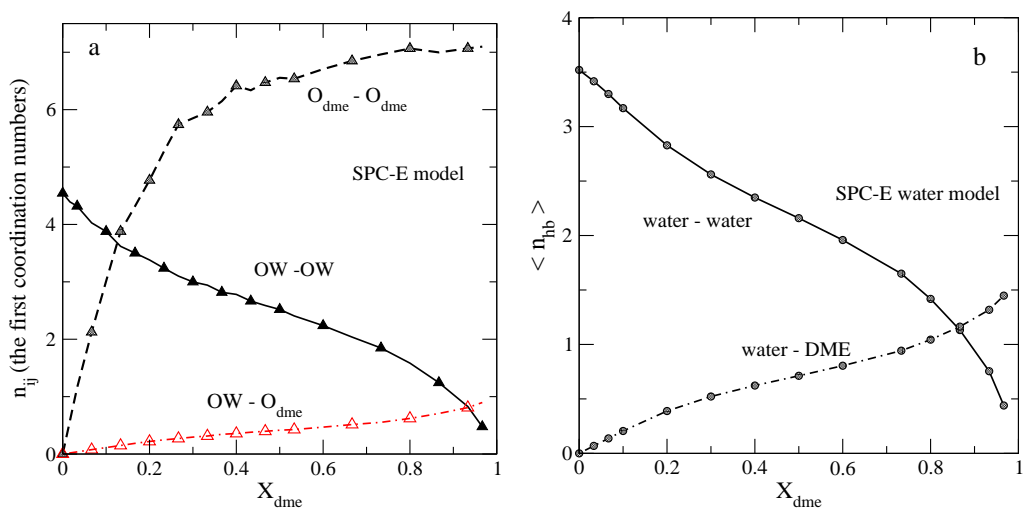

\begin{center}
\includegraphics[width=6.5cm,clip]{fig4a.eps}\quad
\includegraphics[width=6.5cm,clip]{fig4b.eps}
\end{center}
\vspace{-3mm}
\caption{\label{fig6} (Color online) Panel (a): Changes of the first coordination numbers $n_{\text{O}_{\text{dme}}\text{-}\text{OW}}$
and $n_{\text{OW-OW}}$ with mixture composition. Panel~(b): Changes of the average number of hydrogen bonds
between water and DME molecules.}
\end{figure}

The pair distribution functions yield the running coordination numbers by the
relation,
\begin{equation}
 n_i(R)=4\piup\rho_j\int_{0}^{R}g_{ij}(r)r^{2}\rd r,
\end{equation}
where $\rho_j$ is the number density of species $j$. The first coordination number
is obtained  by putting $R=r_{\text{min}}$, i.e., at the first minimum of the
corresponding pair distribution function. The evolution of the first coordination
numbers of species with $X_{\text{dme}}$ is given in figure~\ref{fig6}~(a).  It can be seen that the cross coordination 
number grows smoothly, but remains rather small in the entire composition interval.
The O$_{\text{dme}}$-O$_{\text{dme}}$ coordination number grows with an increasing fraction of organic species, but
does not substantially change in the interval from 0.3 up to DME-rich composition in accordance to
what was discussed for the behaviour of the corresponding pair distribution function. 
Thus, the structure of organic subsystem remains relatively inert to changes of composition. 
The OW-OW  coordination number decreases in magnitude non-monotonously as a result of a decreasing 
fraction of water while $X_{\text{dme}}$ increases, and due to an increasing height of the first maximum of the
OW-OW pair distribution function. A less decline is observed in the interval of composition 
where the first maximum of the pair distribution function increases most drastically.

Next, we would like to comment on the hydrogen bonds network behaviour for the mixtures in question.
The GROMACS software was straightforwardly applied to do that using default distance-angle 
criterion. The average number of hydrogen bonds per water molecule, $\langle n_{\text{HB}}\rangle $,  as a function of 
composition, $X_{\text{dme}}$,
shown in figure~\ref{fig6}~(b), exhibits certain similarities to the behaviour of coordination 
numbers discussed above.
At very low values of $X_{\text{dme}}$, the value for water-water $\langle n_{\text{HB}}\rangle $ is high
serving as an evidence of the hydrogen bonded network between water molecules. 
If DME molecules are introduced into water, the number of their mutual contacts increases, 
as it follows from the behaviour of the pair distribution function OW-O$_{\text{dme}}$.
Consequently, the average number of H-bonds between waters decreases whereas the fraction of
bonds between water molecules and DME oxygens steadily increases. This behaviour is qualitatively similar
to some other water-organic solvent mixtures, see, e.g., figure~8 of our recent study of 
water-DMSO mixtures~\cite{gujt}. In both cases, i.e., the DMSO and DME, no hydrogen bonding is 
feasible in pure organic component.
Most important is that neither the behavior of the coordination numbers as function
of intermolecular distance nor the average number of hydrogen bonds 
provide confirmation that clusters of H-bonded water molecules
can form in DME-water mixtures at high values of $X_{\text{dme}}$. Thus, possible local
heterogeneity of the distribution of water molecules deduced from the behaviour of the pair 
distribution functions does not lead to the formation of big clusters or 
isolated  ``islands'' of water species
distributed in the media of DME molecules at intermediate and high values of $X_{\text{dme}}$.
In other words, miscibility of species is preserved in the entire range of composition.
Still, the formation of strongly ``associated'' species involving groups of water and DME molecules
cannot be discarded.

\begin{figure}[!t]
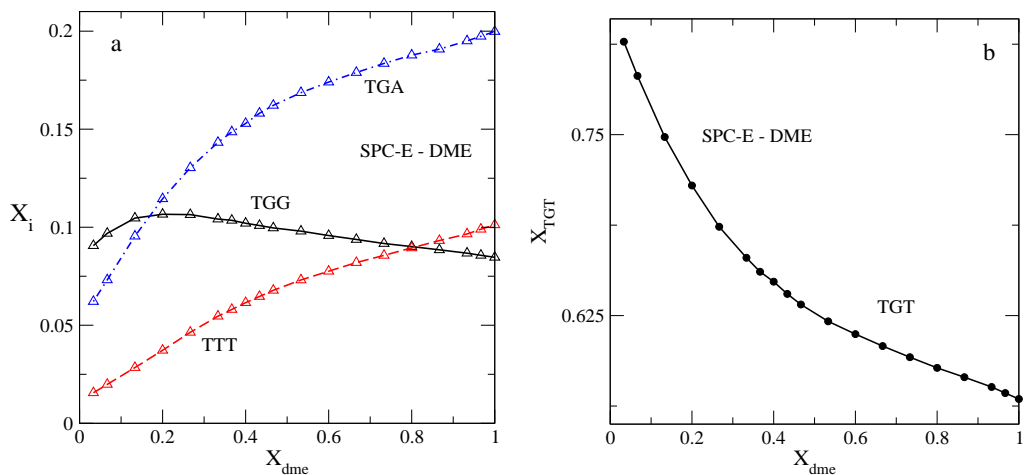

\begin{center}
\includegraphics[width=6.5cm,clip]{fig5a.eps}\quad
\includegraphics[width=6.5cm,clip]{fig5b.eps}
\end{center}
\vspace{-2mm}
\caption{\label{fig7} (Color online) Changes of the fractions of most populated conformations 
of DME molecules with mixture composition.}
\end{figure}

Our final remark in this subsection about the microscopic structure of 
water-DME mixtures concerns the
conformations of DME molecules in mixtures  as functions of the amount of 
the organic co-solvent, $X_{\text{dme}}$ (figure~\ref{fig7}). Those 
were discussed in several recent publications~\cite{sadowski,bedrov,roccatano}. 
We use the same definition of conformers and the same nomenclature $T$, $G$, just $G'$ is
denominated as $A$.
Moreover, average values for conformations population of molecules in liquid DME 
and water-DME mixtures 
are available from Raman spectroscopy~\cite{raman}. It is known that the 
modified TraPPE model in combination with TIP4P-Ew water model leads to 
non-perfect, but in general reasonable description of the conformations 
population upon DME fraction, see, e.g., figure~4 of \cite{sadowski}. 
We were able to reproduce fractions of all conformers reported in~\cite{sadowski},
but failed to reproduce fractions of conformers reported in~\cite{roccatano}.
Accurate description of the population of conformers is of importance on its own
and, moreover, has implications for possible future studies of the effects of solutes
in water-DME solvents. Our results at $T = 298.15$~K, reported in figure~\ref{fig7} exhibit similar trends
as the behaviour reported by Fischer et al. at $T = 318$~K using the TIP4P-Ew water model~\cite{sadowski}. 
Of particular importance is that both the SPC-E and TIP4P-Ew combined with the modified 
TraPPE successfully reproduce experimental observations that should decide which of the conformers are most
populated and how these populations change with $X_{\text{dme}}$.

\subsection{Self-diffusion coefficients and the dielectric constant}

\begin{figure}[!t]
\begin{center}
\includegraphics[width=6.5cm,clip]{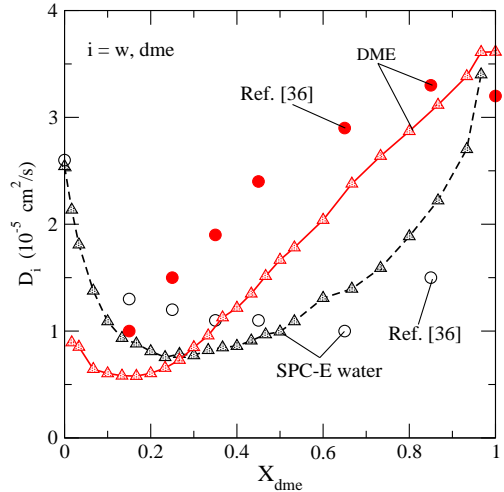}
\end{center}
\vspace{-1mm}
\caption{\label{fig8} (Color online) Composition dependence of the self-diffusion coefficients of species in water-DME mixture from $NPT$ MD simulations.}
\end{figure}

The self-diffusion coefficients of water and DME species in our work  were calculated 
 from the mean-square displacement (MSD) of a particle via Einstein relation,
\begin{equation}
D_i =\frac{1}{6} \lim_{t \rightarrow \infty} \frac{\rd}{\rd t} \langle\vert {\bf 
r}_i(\tau+t)-{\bf r}_i(\tau)\vert ^2\rangle,
\end{equation}
where $i$ refers to water or DME and $\tau$ is the time origin. 
Default settings of GROMACS were used for the separation of the time origins. 
A set of trajectories coming from several consecutive simulations of $10$~ns was
combined to get the entire trajectory not less than 60--70~ns.
The fitting interval then was chosen from $\approx 10$\% to $\approx 50$\% 
of the entire trajectory to obtain  $D_{\text{dme}}$ and $D_\text{w}$.
Moreover, to keep our consciousness calm, we recalculated the 
self-diffusion coefficient for DME molecules
using the center of mass  to yield $D_{\text{dme}}$.
Two sets of results obtained were very close, possibly due to a rather small 
size of DME molecule. 
A set of our results is given in  figure~\ref{fig8}. They concern the SPC-E water model
in conjunction with the modified TraPPE model for DME at $T = 298.15$~K. 
Previous comparisons,
performed in \cite{sadowski} refer to the TIP4P-Ew water model with the
modified TraPPE model at $T = 318$~K. However, only the $D_{\text w}$ values were discussed.
Experimental data at this temperature indicate a minimum of $D_{\text w}$ at $X_{\text{dme}} \approx 0.2$.
Moreover, according to Bedrov et al.~\cite{bedrov2} the experimental value for
$D_{\text{dme}}$ at $X_{\text{dme}} = 1$ is at 3.2. Our calculations yield a correct value for
pure SPC-E water $D_{\text w} = 2.54$ and yield a minimum value of the self-diffusion coefficient
for water species at $X_{\text{dme}} = 0.2333$. On the other hand, we obtained $D_{\text{dme}} \approx 3.6$
for pure DME. In the absence of trustworthy experimental data at $T = 298.15$~K, we believe that
the results both for $D_{\text w}$ and $D_{\text{dme}}$ are at least qualitatively correct. In addition,
we have plotted a rather limited set of points obtained from simulations of the SPC-E water
model and another DME model taken from the table~IIISb of \cite{roccatano}. It is difficult
to explain a peculiar behaviour of $D_{\text{dme}}$ at high values of $X_{\text{dme}}$ (last two points). The
position of a minimum of $D_{\text w}$ also seems to be unsatisfactory. From an
overall shape of the behavior of self-diffusion coefficients of species, one can conclude
that the best  mixing, according to, e.g., the excess molar volume and enthalpy, corresponds
to the minima of self-diffusion coefficients.

Our final remarks in this subsection concern the  behaviour of the static 
dielectric constant, $\varepsilon$,  of mixtures in question. We explore it
as function of the chemical composition. This property was not explored in the previous works on the subject in spite of the experimental data
available.
Usually, the static dielectric constant is calculated from the time-average of 
the fluctuations of the total  dipole moment of the system~\cite{martin}, 
\begin{equation}
\varepsilon=1+\frac{4\piup}{3k_{\text B}TV}(\langle\textbf{M}^2\rangle-\langle\textbf{M}\rangle^2),
\end{equation}
where $k_{\text B}$ is Boltzmann's constant and $V$ is the simulation cell volume. 

\begin{figure}[!t]
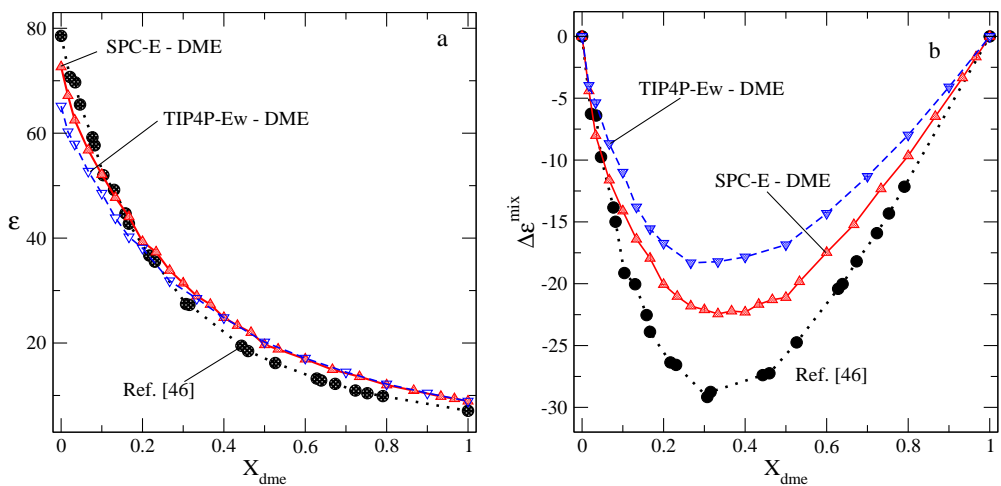

\begin{center}
\includegraphics[width=6.2cm,clip]{fig7a.eps}\quad
\includegraphics[width=6.5cm,clip]{fig7b.eps}
\end{center}
\caption{\label{fig9} (Color online) Panel (a): Composition dependence of the static dielectric constant of
water-DME mixtures. Panel~(b): Excess mixing dielectric constant.}
\end{figure}

The respective curves from simulations for combined water-DME mixtures 
(with the SPC-E and TIP4P-Ew water models) are shown in figure~\ref{fig9}~(a).
General trends of the behaviour of this property show that $\varepsilon $ increases
with decreasing $X_{\text{dme}}$ starting from a low value of the pure substance without
intermolecular hydrogen bonding (DME) to a much higher value corresponding to pure water.
As it follows from the comparison of the simulation results and experimental data~\cite{renard},
both  models underestimate the values for $\varepsilon$ on the water-rich 
side and slightly overestimate the dielectric constant on the DME-rich side. 
The static dielectric constant for pure DME due to our
calculations is $\varepsilon \approx 8.9$ 
whereas the experiment yields 7.08~\cite{renard}.  
Nevertheless, the discrepancies  between the simulation
results  and experiment are not big in the entire interval of composition changes. 
As expected, the SPC-E model combined with
the modified TraPPE yields slightly better results in the water-rich interval
of composition.

A more sensitive test is provided by a comparison of the excess 
dielectric constant [figure~\ref{fig9}~(b)], 
$\Delta \varepsilon_{\text{mix}} = \varepsilon-(1-X_{\text{dme}})\varepsilon_{\text{water}}-X_{\text{dme}}\varepsilon_{\text{dme}}$,
 with the experimental predictions~\cite{renard}. 
Experimental points indicate a negative deviation from
perfection in the entire composition range. Interestingly, this behaviour is contrary
to what follows for water-DMSO liquid mixtures~\cite{gujt}.
Maximal deviation from the ideal type behaviour
reported from the experimental measurements for the system in question is at $X_{\text{dme}} = 0.3$. 
The simulation results
reproduce the position of minimum successfully. However, the magnitude of deviation
is underestimated from the simulated models. The SPC-E with a modified TraPPE model is closer
to the experimental values~\cite{renard},  compared with the TIP4P-Ew. 
Another set of experimental data concerning $\Delta \varepsilon_{\text{mix}}$ dependence
on composition was reported in~\cite{prabhu}. Again, the minimum value of the
excess mixing static dielectric constant was observed at $X_{\text{dme}} \approx  0.3$. However,
the value at minimum is around 25, showing close agreement of the SPC-E-TraPPE
model predictions with these experimental observations. 
It is worth mentioning that a negative deviation
of the excess dielectric constant for mixtures surely results from the presence of
``associated'' mixed species of water and DME molecules with favourable correlation of the
dipole moments of molecules, i.e., their symmetrical orientation
contributing to polarizability,  and certain amount of cross hydrogen bonds. 
Moreover, the lifetime of these 
species or complexes seems to be such that it can be detected by the dielectric constant measurements.

\section{Summary and conclusions}

The mixtures explored in this work are one of the examples from the class of systems 
composed of water and organic co-solvent. They are of  much importance in laboratory studies
with possible applications in chemistry and bio-related areas.

We have performed an extensive set of molecular dynamic simulations in 
the isobaric-isothermal  ensemble to study the density, mixing properties and the microscopic 
structure of  water-DME mixtures in the entire range of a
solvent composition. The self-diffusion coefficients of species and the 
dielectric constant were calculated as well. 
All the simulations were performed at room temperature and ambient pressure, 
1~bar.  Two water models (SPC-E and TIP4P-Ew), combined with the 
modified TraPPE model for DME~\cite{sadowski},
were studied; we consider this as a first step of systematic studies
of mixtures of water and DME at different thermodynamic
states using nonpolarizable models.

From a comparison with the available experimental data for different properties and
with the results of other authors on this and related systems, 
we can conclude that the predictions
obtained are qualitatively correct and give a physically sound picture of the properties explored.
We observed that practically all the properties investigated are sensitive to the
composition of water-organic liquid solvent.

The principal conclusions of the present study can be summed up as follows.
We explored the evolution of the microscopic
structure in terms of the pair distribution functions together with
the first coordination numbers of species. In this respect, the simulation results
evidence that the structure of the subsystem of DME species is much more
inert or much less sensitive to the composition, in comparison with the structure
of an aqueous subsystem. The pair distribution functions for water species evidence 
that a heterogeneous density distribution at local scale can develop upon adding the DME
molecules. These trends of changes of the microscopic structure bear a similarity to
some other water-organic co-solvent mixtures, see, e.g.,~\cite{kovacs,mountain}.

However, from the behaviour of the coordination numbers in the system explored, we have not 
found that  water clusters of significant size can be formed. 
The cross correlations seem to be not very strong as it follows
from the corresponding coordination number, but mixing is well pronounced.
Thus, the ``associated'' species involving water and DME molecules or their groups
are formed in the system, with or without hydrogen bonds.
Statistical features of the hydrogen bonding
in water  and of water-organic co-solvent bonds do not show a peculiar
behaviour, in comparison to a qualitatively similar  mixture of water with DMSO~\cite{gujt}.
On the other hand, the analyses of the most popular conformations and their fractions
as function of $X_{\text{dme}}$ at room temperature are in accordance with the recently
reported results of~\cite{sadowski}.

Dynamic properties have been studied from the mean square displacements and are given
in terms of  self-diffusion coefficients of species, $D_{\text w}$ and $D_{\text{dme}}$. Both
of them exhibit a minimum in the interval of composition that corresponds to 
most ``packed'' structures according to the behaviour of mixing volume. The values
of self-diffusion coefficients of species for pure components are reasonably well
described. Concerning the dependence of the static dielectric constant on composition, 
we observe that its excess is better described while using the SPC-E water model in
comparison with the TIP4P-Ew model. It would be of interest to relate the behaviour of
the static dielectric constant on composition with refractive index and viscosity data. 
However, this would require an additional computer simulation work. 

At the present stage of the development, the missing elements worth a more detailed investigation 
include the application of other, more elaborate, force fields for the DME molecules 
for the description of water-DME mixtures. Wider insights into the behaviour of 
dynamic and dielectric properties by exploration, e.g., the relaxation times, 
hydrogen-bonds lifetime and complex dielectric constant,  would be desirable. 
Unfortunately, the structure factors of the mixtures in question from the
diffraction experiments are unavailable. Hence, for the moment  we are unable to perform
a detailed study of the microscopic structure along the lines proposed in \cite{pusztai}
and recently applied for water-methanol mixtures \cite{galicia1}.

Our interest into the properties of water-DME mixtures has been principally inspired  
by possible and challenging extensions. One of them, rather straightforward, is the
application of molecular dynamics simulations to explore  longer
polyoxyethylene oligomers that exhibit more complex and richer properties. 
Another challenging line of research is to explore the solutions with ionic
solutes or complex molecules in water-DME mixtures in close similarity to the
previous combined experimental and simulation studies, in the spirit of~\cite{ewa1,ewa2,ewa3,ewa4} or of 
e.g.~\cite{takamuku2,takamuku3,takamuku4,takamuku5,taha}. In this latter group
of works, the phase diagrams of solute-combined water-organic solvent systems
were described using only experimental tools. 
Thus, the proposed description of possible phase separation mechanisms seems to be vague.
Actually, the problem
of ionic solutes in water-organic solvent mixtures has been successfully considered
only for the case of two solvent components being immiscible, see, e.g.,~\cite{holmberg}.
These problems are now under study in our laboratory. 

\section*{Acknowledgements}
O.P. is grateful to D. Vazquez and M. Aguilar for technical support of this work
at the Institute of Chemistry of the UNAM.

\ukrainianpart
\title{Комп'ютерне моделювання методом молекулярної динаміки в ізобарично-ізотермічному ансамблі властивостей модельних сумішей 
вода-1,2-диметоксиетан
}

\author{Ю. Гуйт\refaddr{label1}, Г.  Домінгез\refaddr{label2},
С. Соколовскі\refaddr{label3},
O. Пізіо\refaddr{label2}}

\addresses{
  \addr{label1} Кафедра теоретичної фізики, хімічний факультет, університет Дуйсбург-Ессен,
D-45141 Ессен, Німеччина
\addr{label2} Інститут матеріалознавства, Нацiональний автономний унiверситет м. Мехiко, Мехiко, Мексика
\addr{label3} Відділ моделювання фізико-хімічних процесів, університет Марії-Склодовської Кюрі, Люблін, Польща
}

\makeukrtitle

\begin{abstract}
Для того, щоб дослідити широкий 
набір властивостей модельних сумішей вода-1,2-диметоксиетан (DME) в залежності від концентрації,  
проведено комп'ютерне моделювання методом молекулярної динаміки в ізобарично-ізотермічному ансамблі. Для води застосовано моделі 
SPC-E і TIP4P-Ew, а для DME --- модифіковану  модель TraPPE. Нашим основним завданням було дослідити тенденцію поведінки  
структурних властивостей в термінах радіальних функцій розподілу, координаційних чисел та чисел водневих зв'язків 
між молекулами різних сортів, а також конформації молекул DME.
Вивчено термодинамічні властивості, такі як густина, молярний об'єм, ентальпія змішування і питома теплоємність при постійному тиску.
Накінець, обчислено і проаналізовано коефіцієнти самодифузії сортів і діелектричну сталу системи.

\keywords  суміші вода-DME, термодинамічні властивості, коефіцієнти самодифузії, діелектрична стала, молекулярна динаміка

\end{abstract}

\end{document}